\documentclass[twocolumn,aps,superscriptaddress,showpacs]{revtex4}
\usepackage{amssymb}
\usepackage{amsmath}
\usepackage{graphicx}
\usepackage[normalem]{ulem}
\usepackage[dvips]{color}

\renewcommand\sout{\bgroup \color{red} \ULdepth=-.5ex \ULset}

\begin{document}

\title{Effect of the momentum dependence of nuclear symmetry potential on the
transverse and elliptic flows}
\author{Lei Zhang}
\affiliation{School of Information Engineering, Hangzhou Dianzi
University, Hangzhou 310018, China}
\author{Yuan Gao\footnote{Corresponding author: gaoyuan@impcas.ac.cn}}
\affiliation{School of Information Engineering, Hangzhou Dianzi
University, Hangzhou 310018, China} \affiliation{School of Nuclear
Science and Technology, Lanzhou University, Lanzhou 730000, China}
\author{Yun Du}
\affiliation{School of Information Engineering, Hangzhou Dianzi
University, Hangzhou 310018, China}
\author{Guang-Hua Zuo}
\affiliation{School of Information Engineering, Hangzhou Dianzi
University, Hangzhou 310018, China}
\author{Gao-Chan Yong}
\affiliation{Institute of Modern Physics, Chinese Academy of
Sciences, Lanzhou 730000, China}

\begin{abstract}

In the framework of the isospin-dependent
Boltzmann-Uehling-Uhlenbeck transport model, effect of the
momentum dependence of nuclear symmetry potential on nuclear
transverse and elliptic flows in the neutron-rich reaction
$^{132}$Sn+$^{124}$Sn at a beam energy of $400$ MeV/nucleon is
studied. We find that the momentum dependence of nuclear symmetry
potential affects the rapidity distribution of the free neutron to
proton ratio, the neutron and the proton transverse flows as a
function of rapidity. The momentum dependence of nuclear symmetry
potential affects the neutron-proton differential transverse flow
more evidently than the difference of neutron and proton
transverse flows as well as the difference of proton and neutron
elliptic flows. It is thus better to probe the symmetry energy by
using the difference of neutron and proton flows since the
momentum dependence of nuclear symmetry potential is still an open
question. And it is better to probe the momentum dependence of
nuclear symmetry potential by using the neutron-proton
differential transverse flow the rapidity distribution of the free
neutron to proton ratio.

\end{abstract}

 \maketitle

\section{Introduction}

Nowadays the equation of state (EOS) of isospin symmetric nuclear
matter is now relatively well determined mainly by studying
collective flows in heavy-ion collisions and nuclear giant
monopole resonances \cite{pd02,youngblood99}. The major remaining
uncertainty about the EOS of symmetric nuclear matter is due to
our poor knowledge about the density dependence of the nuclear
symmetry energy \cite{LCK08,Bar05,pd02,pie04,colo04}, which is
crucial for understanding many interesting issues in both nuclear
physics and astrophysics \cite{Bro00,Sum94,Lat04,Ste05a}. And it
is also crucial in connection with the structure of neutron stars
and the dynamical evolution of proto-neutron stars \cite{mk94}.
Nowadays considerable progress has been made recently in
determining the density dependence of the nuclear symmetry energy
around the normal nuclear matter density. However, much more work
is still needed to probe the high-density behavior of the nuclear
symmetry energy. Currently, to pin down the symmetry energy, the
National Superconducting Cyclotron Laboratory (NSCL) at Michigan
State University, the Gesellschaft fuer Schwerionenforschung (GSI)
at Darmstadt, the Rikagaku Kenkyusho (RIKEN, the Institute of
Physical and Chemical Research) of Japan, and the Cooler Storage
Ring (CSR) in Lanzhou are planning to do related experiments to
probe the symmetry energy.

The neutron-proton differential transverse flow and the difference
of neutron and proton flows are both sensitive to the symmetry
energy \cite{LCK08,Bar05}, but the used transport models always
adopt different momentum dependent interactions among nucleons.
The importance of the momentum dependence of nuclear symmetry
potential on the two kinds of nuclear flows was seldom mentioned.
Considering the momentum dependence of nuclear symmetry potential
is quite controversial \cite{chen11}, in the framework of the
isospin-dependent Boltzmann-Uehling- Uhlenbeck transport model, we
find that, besides the ratio of $\pi^{-}/\pi^{+}$ \cite{gao11},
the momentum dependence of nuclear symmetry potential affects the
neutron-proton differential transverse flow more evidently than
the difference of neutron and proton transverse flows as well as
the difference of proton and neutron elliptic flows. It is thus
better to probe the symmetry energy by using the difference of
nucleonic transverse (or elliptic) flows.

\section{THE IBUU04 TRANSPORT MODEL}

The present study is based on the IBUU04 transport model
\cite{LCK08}. The initial neutron and proton density distributions
of the projectile and target are obtained by using the
relativistic mean field theory. Although nuclear flow may be
affected by the initializations of colliding nuclei \cite{init11},
our main results in the present studies are not sensitive to the
initializations of target and projectile nuclei. The experimental
free-space nucleon-nucleon (NN) scattering cross sections and the
in-medium NN cross sections can be used optionally. In the present
work, we did not use the isospin-dependent in-medium NN elastic
cross sections from the scaling model according to nucleon
effective masses \cite{li05,epj11}, although it is similar to the
Brueckner approach calculations \cite{ligq93,fuchs01,zhf0710}.
This is because the momentum-independent symmetry potential (MID)
can not use the isospin-dependent in-medium NN elastic cross
sections from the scaling model according to nucleon effective
masses. Thus all the studies here optionally use the experimental
free-space nucleon-nucleon (NN) scattering cross sections for
consistence. For the inelastic cross sections people use the
experimental data from free space NN collisions since the
in-medium inelastic NN cross sections are still very much
controversial. The total and differential cross sections for all
other particles are taken either from experimental data or
obtained by using the detailed balance formula. In the model,
besides nucleons, $\Delta $ and $N^{\ast }$ resonances as well as
pions and their isospin-dependent dynamics are included. The
isospin dependent phase-space distribution functions of the
particles involved are solved by using the test-particle method
numerically. The isospin-dependence of Pauli blockings for
fermions is also considered. The momentum-dependent single nucleon
potential (MDI) adopted here is \cite{das03}%
\begin{eqnarray}
U(\rho ,\delta ,\mathbf{p},\tau ) &=&A_{u}(x)\frac{\rho _{\tau ^{\prime }}}{%
\rho _{0}}+A_{l}(x)\frac{\rho _{\tau }}{\rho _{0}}  \nonumber \\
&&+B(\frac{\rho }{\rho _{0}})^{\sigma }(1-x\delta ^{2})-8x\tau \frac{B}{%
\sigma +1}\frac{\rho ^{\sigma -1}}{\rho _{0}^{\sigma }}\delta \rho
_{\tau
^{\prime }}  \nonumber \\
&&+\frac{2C_{\tau ,\tau }}{\rho _{0}}\int d^{3}\mathbf{p}^{\prime }\frac{%
f_{\tau }(\mathbf{r},\mathbf{p}^{\prime
})}{1+(\mathbf{p}-\mathbf{p}^{\prime
})^{2}/\Lambda ^{2}}  \nonumber \\
&&+\frac{2C_{\tau ,\tau ^{\prime }}}{\rho _{0}}\int d^{3}\mathbf{p}^{\prime }%
\frac{f_{\tau ^{\prime }}(\mathbf{r},\mathbf{p}^{\prime })}{1+(\mathbf{p}-%
\mathbf{p}^{\prime })^{2}/\Lambda ^{2}}.  \label{potential}
\end{eqnarray}%
In the above equation, $\delta =(\rho _{n}-\rho _{p})/(\rho
_{n}+\rho _{p})$ is the isospin asymmetry parameter, $\rho =\rho
_{n}+\rho _{p}$ is the baryon density and $\rho _{n},\rho _{p}$
are the neutron and proton densities, respectively. $\tau
=1/2(-1/2)$ for neutron (proton) and $\tau \neq \tau ^{\prime }$,
$\sigma =4/3$, $f_{\tau }(\mathbf{r},\mathbf{p})$ is the
phase-space distribution function at coordinate $\mathbf{r}$ and
momentum $\mathbf{p}$. The parameters $A_{u}(x),A_{l}(x),B,C_{\tau
,\tau }$, $C_{\tau ,\tau ^{\prime }}$ and $\Lambda $ were set by
reproducing the momentum-dependent potential $U(\rho ,\delta
,\mathbf{p},\tau )$ predicted by the Gogny Hartree-Fock and/or the
Brueckner-Hartree-Fock calculations, the saturation properties of
symmetric nuclear matter and the symmetry
energy of about $32$ MeV at normal nuclear matter density $\rho _{0}=0.16$ fm%
$^{-3}$. The propagations of nucleon are according to Hamilton's
equations
\begin{eqnarray}\label{heq}
dp_{i}/dt&=&-\nabla_{r} U(r_{i})+q_{i}\vec{E},\nonumber\\
dr_{i}/dt&=&p_{i}/\sqrt{m^{2}+p_{i}^{2}}+\nabla_{p}
U(r_{i},p_{i}),
\end{eqnarray}
where $q_{i}$ is the charge of particle, $\vec{E}$ is the Coulomb
field of particle felt. The incompressibility of symmetric nuclear
matter at normal density is set to be $211$ MeV. According to
essentially all microscopic model calculations, the EOS for
isospin asymmetric nuclear matter can be expressed as
\begin{equation}
E(\rho ,\delta )=E(\rho ,0)+E_{\text{sym}}(\rho )\delta ^{2}+\mathcal{O}%
(\delta ^{4}),
\end{equation}%
where $E(\rho ,0)$ is the energy per nucleon of symmetric nuclear
matter, and $E_{\text{sym}}(\rho )$ is the nuclear symmetry
energy. With the single particle potential $U(\rho ,\delta
,\mathbf{p},\tau )$, for a given
value $x$, one can readily calculate the symmetry energy $E_{\text{sym}%
}(\rho )$ as a function of density. Because the purpose of present
studies is just to see how large the effect of momentum dependence
of nuclear symmetry potential on the transverse and elliptic
flows, we let the variable $x$ be $1$, since the IBUU04 model
gives a super-soft symmetry energy at higher densities
\cite{xiao09}. In fact, behavior of nuclear symmetry energy at
supra-densities is still in controversy. The main characteristic
of the present single particle is the momentum dependence of
nuclear symmetry potential, which has evident effect on energetic
free $n/p$ ratio in heavy-ion collisions \cite{IBUU04}. In the
present studies, to show the effects of the momentum dependence of
symmtry potential, we kept the isoscalar potential fixed while
changing the symmetry potential from the momentum dependent
symmetry potential to the momentum independent symmetry potential
and keep the symmetry energy fixed \cite{IBUU04}.
\begin{figure}[th]
\begin{center}
\includegraphics[width=0.5\textwidth]{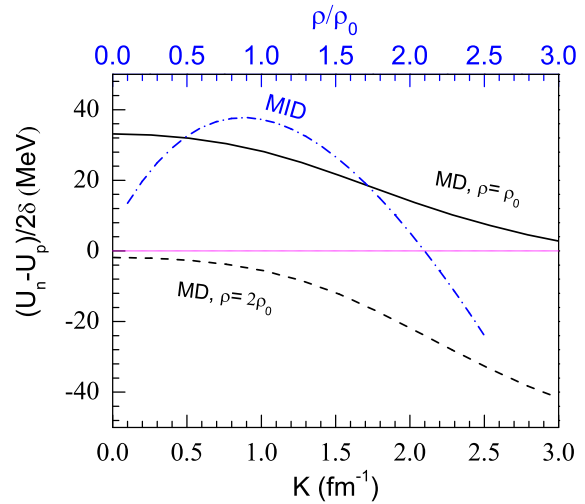}
\end{center}
\caption{(Color online) The momentum dependent symmetry potential
(MD) as a function of momentum and the momentum independent
symmetry potential (MID) as a function of density (upper
x-coordinate) from the Gogny interaction \cite{IBUU04}.}
\label{symp}
\end{figure}
Fig.~\ref{symp} shows the momentum dependent symmetry potential
(MD) and the momentum independent symmetry potential (MID) from
the Gogny interaction. We can see that at higher densities
strength of the momentum independent symmetry potential (MID) is
always larger than that of the momentum dependent symmetry
potential. In the following, we give our results of the momentum
dependence of nuclear symmetry potential on nuclear transverse and
elliptic flows at higher densities (The density reached in the
$^{132}$Sn+$^{124}$Sn at a beam energy of 400 MeV/nucleon is about
2 times saturation density \cite{yong06}).

\section{Results and discussions}

\begin{figure}[th]
\begin{center}
\includegraphics[width=0.5\textwidth]{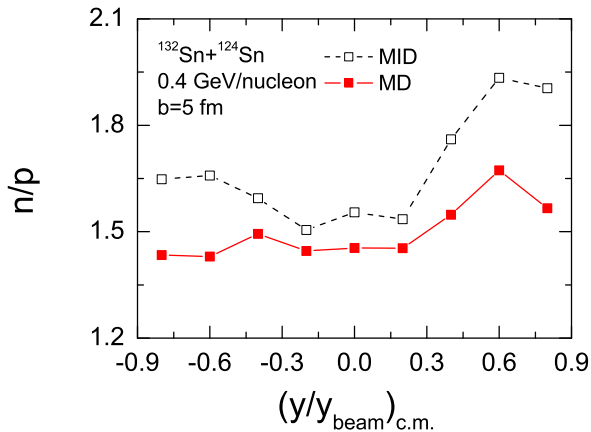}
\end{center}
\caption{(Color online) Rapidity distribution of free neutron to
proton ratio n/p in the reaction $^{132}$Sn+$^{124}$Sn at a beam
energy of 400 MeV/nucleon and an impact parameter of 5 fm with and
without momentum dependence of nuclear symmetry potential, signed
with MD and MID, respectively.} \label{yrnp}
\end{figure}
We first study the neutron to proton ratio n/p of free nucleons as
a function of rapidity as shown in Fig.~\ref{yrnp}. It is seen the
case with momentum dependence of nuclear symmetry potential causes
lower neutron to proton ratio, whereas the case without momentum
dependence of nuclear symmetry potential causes higher neutron to
proton ratio, especially for nucleons at large rapidities. Because
the momentum dependence of nuclear symmetry potential decreases
the strength of nuclear symmetry potential at high densities or
high nucleonic momenta as shown in Fig.~\ref{symp}. The small
symmetry potential decreases the free neutron to proton ratio
\cite{IBUU04}. From this plot, we can see that while using the
neutron to proton ratio n/p of free nucleons to probe the symmetry
energy, one should keep in mind that this observable is sensitive
to the the momentum dependence of nuclear symmetry potential used,
and thus may cause uncertainties while compared with the
experimental data.

\begin{figure}[th]
\begin{center}
\includegraphics[width=0.5\textwidth]{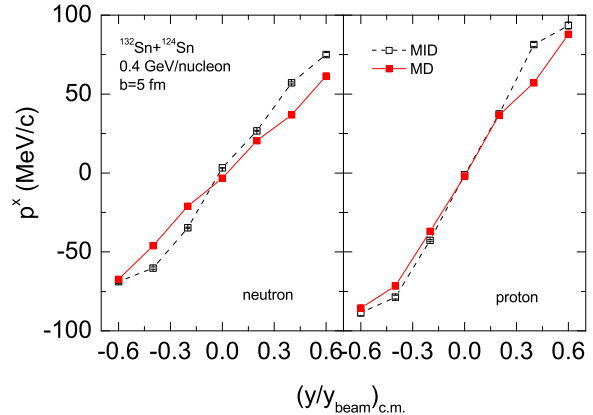}
\end{center}
\caption{(Color online) Neutron and proton transverse flows
analysis in the reaction $^{132}$Sn+$^{124}$Sn at a beam energy of
400 MeV/nucleon and an impact parameter of 5 fm with and without
momentum dependence of nuclear symmetry potential, signed with MD
and MID, respectively.} \label{sflow}
\end{figure}
Fig.~\ref{sflow} shows neutron and proton transverse flows
analysis in the reaction $^{132}$Sn+$^{124}$Sn at a beam energy of
400 MeV/nucleon and an impact parameter of 5 fm with and without
momentum dependence of nuclear symmetry potential. We can see that
without (with) momentum dependence of nuclear symmetry potential,
the strength of the nucleon flow (especially neutron flow)
increases (decreases). This is understandable since the momentum
independence of nuclear symmetry potential overall increases the
strength of nuclear symmetry potential as shown in
Fig.~\ref{symp}, thus neutrons are repelled more strongly than the
case with the momentum dependent symmetry potential. From
Fig.~\ref{sflow} we can also see that effect of the momentum
dependence of nuclear symmetry potential on the proton flow is
less evident owing to the Coulomb potential added on protons. The
Coulomb potential (always repulsive for protons) decreases the
strength of symmetry potential (attractive for protons at high
densities) added on protons.

\begin{figure}[th]
\begin{center}
\includegraphics[width=0.5\textwidth]{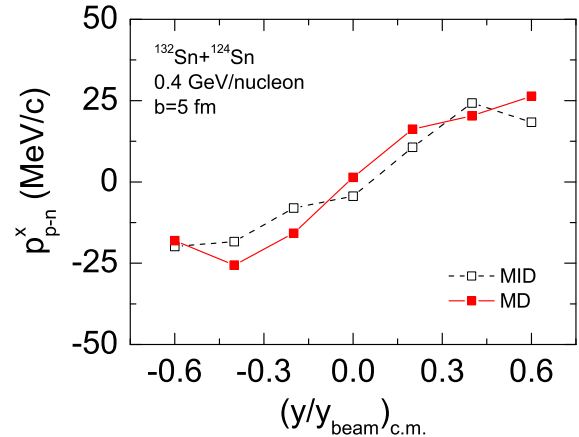}
\end{center}
\caption{(Color online) Difference of proton and neutron
transverse flows analysis in the reaction $^{132}$Sn+$^{124}$Sn at
a beam energy of 400 MeV/nucleon and an impact parameter of 5 fm
with and without momentum dependence of nuclear symmetry
potential, signed with MD and MID, respectively.} \label{dflow}
\end{figure}
The difference of proton and neutron transverse flows has been
studied previously and was shown to be sensitive to the symmetry
energy \cite{Bar05}. In order to show if the difference of proton
and neutron transverse flows is sensitive to the momentum
dependence of nuclear symmetry potential, we plot
Fig.~\ref{dflow}, the difference of proton and neutron transverse
flows analysis in the reaction $^{132}$Sn+$^{124}$Sn at a beam
energy of 400 MeV/nucleon and an impact parameter of 5 fm with and
without momentum dependence of nuclear symmetry potential. From
this plot, we see that the difference of proton and neutron
transverse flows is insensitive to the momentum dependence of
nuclear symmetry potential. It is thus better to be used to probe
the symmetry energy since the the momentum dependence of nuclear
symmetry potential is still an open question \cite{chen11}.

\begin{figure}[th]
\begin{center}
\includegraphics[width=0.5\textwidth]{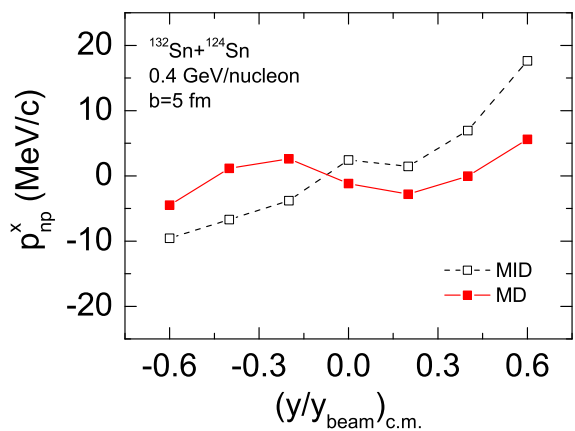}
\end{center}
\caption{(Color online) Neutron proton differential transverse
flow analysis in the reaction $^{132}$Sn+$^{124}$Sn at a beam
energy of 400 MeV/nucleon and an impact parameter of 5 fm with and
without momentum dependence of nuclear symmetry potential, signed
with MD and MID, respectively.} \label{cflow}
\end{figure}
The neutron proton differential transverse flow also has been
studied extensively and was shown to be sensitive to the symmetry
energy \cite{LCK08,IBUU04}. To see if neutron proton differential
transverse flow is sensitive to the momentum dependence of nuclear
symmetry potential, we give Fig.~\ref{cflow}, the neutron proton
differential transverse flow analysis in the reaction
$^{132}$Sn+$^{124}$Sn at a beam energy of 400 MeV/nucleon and an
impact parameter of 5 fm with and without momentum dependence of
nuclear symmetry potential. From this plot, we see that the
neutron proton differential transverse flow is sensitive to the
momentum dependence of nuclear symmetry potential as shown in
\cite{IBUU04}.

The neutron-proton differential transverse flow is defined as
\cite{yong06,yong09}
\begin{eqnarray}
F_{n-p}^{x}(y) &\equiv
&\frac{1}{N(y)}\sum_{i=1}^{N(y)}p_{i}^{x}(y)w_{i}
\nonumber \\
&=&\frac{N_{n}(y)}{N(y)}\langle p_{n}^{x}(y)\rangle -\frac{N_{p}(y)}{N(y)}%
\langle p_{p}^{x}(y)\rangle  \label{npflow}
\end{eqnarray}%
where $N(y)$, $N_{n}(y)$ and $N_{p}(y)$ are the number of free
nucleons, neutrons and protons, respectively, at rapidity $y$;
$p_{i}^{x}(y)$ is the transverse momentum of the free nucleon at
rapidity $y$; $w_{i}=1$ $(-1)$ for neutrons (protons); and
$\langle p_{n}^{x}(y)\rangle $ and $\langle p_{p}^{x}(y)\rangle $
are respectively the average transverse momenta of neutrons and
protons at rapidity $y$. One can see from Eq. (\ref{npflow}) that
the constructed neutron-proton differential transverse flow
depends not only on the proton flow and neutron flow but also on
their relative multiplicities. Therefore the neutron-proton
differential transverse flow is not simply the difference of the
neutron and proton transverse flows, it in fact depends also on
the isospin fractionation at the rapidity $y$. If neutrons and
protons have the same average transverse momentum in the reaction
plane but different multiplicities in each rapidity bin, i.e.,
$\langle p_{n}^{x}(y)\rangle =\langle p_{p}^{x}(y)\rangle =\langle
p^{x}(y)\rangle $, and $N_{n}(y)\neq N_{p}(y)$, then Eq.
(\ref{npflow}) is reduced to
\begin{equation}
F_{n-p}^{x}(y)=\frac{N_{n}(y)-N_{p}(y)}{N(y)}\langle
p^{x}(y)\rangle =\delta (y)\cdot \langle p^{x}(y)\rangle,
\end{equation}%
reflecting effects of the isospin fractionation. On the other
hand, if neutrons and protons have the same multiplicity but
different average transverse momenta, i.e., $N_{n}(y)=N_{p}(y)$
but $\langle
p_{n}^{x}(y)\rangle \neq \langle p_{p}^{x}(y)\rangle $, then Eq. (\ref%
{npflow}) is reduced to%
\begin{equation}
F_{n-p}^{x}(y)=\frac{1}{2}(\langle p_{n}^{x}(y)\rangle -\langle
p_{p}^{x}(y)\rangle ).
\end{equation}%
In this case it reflects directly the difference of the neutron
and proton transverse flows. Because the effect of the momentum
dependence of nuclear symmetry potential for neutron to proton
ratio is very large (shown in Fig.~\ref{yrnp}) and the difference
of the proton and neutron transverse flows has no such
combination, we see larger effect of the momentum dependence of
nuclear symmetry potential on the neutron proton differential
transverse flow than the difference of proton and neutron
transverse flows. It is thus better to probe the symmetry energy
by using the difference of neutron and proton flows since the
momentum dependence of nuclear symmetry potential is still an open
question \cite{chen11}.

\begin{figure}[th]
\begin{center}
\includegraphics[width=0.5\textwidth]{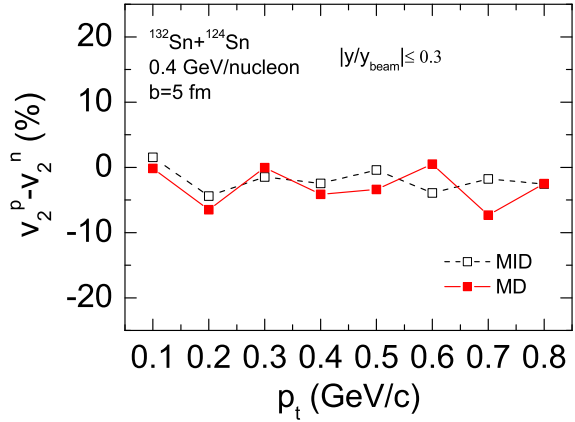}
\end{center}
\caption{(Color online) Difference of proton elliptic flow and
neutron elliptic flow analysis in the reaction
$^{132}$Sn+$^{124}$Sn at a beam energy of 400 MeV/nucleon and an
impact parameter of 5 fm with and without momentum dependence of
nuclear symmetry potential, signed with MD and MID, respectively.}
\label{eflow}
\end{figure}
The elliptic flow $v_{2}(y,p_t)$, which
is derived as the second coefficient from a Fourier expansion of
the azimuthal distribution
$N(\phi,y,p_{t})=v_{0}(1+v_{1}cos(\phi)+2v_{2}cos(2\phi))$, can be
expressed as
\begin{equation}
 v_2=<\frac{p^2_x-p^2_y}{p^2_t}>,
 \end{equation}
where $p_t=\sqrt{p^2_x+p^2_y}$ is the transverse momentum
\cite{ditorof,ditorof2}. The difference of proton and neutron
elliptic flows is also shown to be sensitive to the symmetry
energy \cite{ditorof,ditorof2}. It is thus also interesting to see
if the difference of proton elliptic flow and neutron elliptic
flow is sensitive to the momentum dependence of nuclear symmetry
potential. Fig.~\ref{eflow} shows the difference of proton
elliptic flow and neutron elliptic flow analysis in the reaction
$^{132}$Sn+$^{124}$Sn at a beam energy of 400 MeV/nucleon and an
impact parameter of 5 fm with and without momentum dependence of
nuclear symmetry potential. Again, we see that the difference of
proton elliptic flow and neutron elliptic flow is not sensitive to
the symmetry energy.

\section{Conclusions}

Based on the IBUU04 transport model, effect of the momentum
dependence of nuclear symmetry potential on nuclear transverse and
elliptic flows in the neutron-rich reaction $^{132}$Sn+$^{124}$Sn
at a beam energy of $400$ MeV/nucleon is studied. It is found that
the momentum dependence of nuclear symmetry potential affects the
rapidity distribution of the free neutron to proton ratio, neutron
flow and proton flow as a function of rapidity. The momentum
dependence of nuclear symmetry potential affects neutron-proton
differential transverse flow more evidently than the difference of
neutron and proton transverse flows as well as the difference of
proton elliptic flow and neutron elliptic flow. Therefore it is
better to probe the symmetry energy by using the difference of
neutron flow and proton flow since the momentum dependence of
nuclear symmetry potential is still an open question. And it is
better to probe the momentum dependence of nuclear symmetry
potential by using the neutron-proton differential transverse flow
and the rapidity distribution of the free neutron to proton ratio.

\section*{Acknowledgments}

The author Y. Gao thanks Prof. Bao-An Li for providing the code
and useful guidance while he stayed at Institute of Modern
Physics, Chinese Academy of Sciences and thanks Prof. Wei Zuo for
helpful discussions. The work is supported by the National Natural
Science Foundation of China (10975064, 11175074, 11175219 and
11105035) and the Zhejiang Provincial Natural Science Foundation
(Y6110644).

\end{document}